\documentclass[aps,twocolumn,superscriptaddress,footinbib,nobalancelastpage,prx]{revtex4-2}
\usepackage{amsmath,amssymb}
\usepackage[colorlinks=true,citecolor=blue]{hyperref}
\usepackage{graphicx}
\usepackage{amsmath}
\usepackage{slashed}
\usepackage{physics}

\usepackage[table,xcdraw,dvipsnames]{xcolor}
\usepackage[caption=false]{subfig}

\newcommand{\be}{\begin{equation}}
\newcommand{\ee}{\end{equation}}

\begin{document}

\title{$Q$-Deformed Rainbows: a Universal Simulator of Free Entanglement Spectra}

\author{Lucy Byles}
\email{pylsb@leeds.ac.uk}
\affiliation{School of Physics and Astronomy, University of Leeds, Leeds, LS2 9JT, United Kingdom}
\author{Germ\' an Sierra}
\affiliation{Instituto de F\'isica Te\'orica UAM/CSIC, Universidad Aut\'onoma de Madrid, Madrid, 28049, Spain}
\author{Jiannis K. Pachos}
\affiliation{School of Physics and Astronomy, University of Leeds, Leeds, LS2 9JT, United Kingdom}

\begin{abstract}

The behaviour of correlations across a bipartition is an indispensable tool in diagnosing quantum phases of matter. Here we present a spin chain with position-dependent XX couplings and magnetic fields, that can reproduce arbitrary structure of free fermion correlations across a bipartition. In particular, by choosing appropriately the strength of the magnetic fields we can obtain any single particle energies of the entanglement spectrum with high fidelity. The resulting ground state can be elegantly formulated in terms of $q$-deformed singlets. To demonstrate the versatility of our method we consider certain examples, such as a system with homogeneous correlations and a system with correlations that follow a prime number decomposition. Hence, our entanglement simulator can be easily employed for the generation of arbitrary entanglement spectra with possible applications in quantum technologies and condensed matter physics.

\end{abstract}

\maketitle

\section{Introduction}

Entanglement lies at the heart of the disparity between classical and quantum mechanics. As such it has long been at the forefront of both theoretical \cite{einstein1935can, shimony1993conceptual} and experimental \cite{shadbolt2014testing, tang2022some} investigations into the foundations of quantum mechanics. More recently, entanglement has gained renewed interest with the development of quantum information theory \cite{nielsen2002quantum, jozsa2004illustrating}. In this framework, quantum entanglement is viewed as a valuable resource \cite{jozsa1997entanglement}, with several quantum protocols, such as teleportation \cite{bennett1993teleporting}, able to be realised exclusively with the use of entangled states. This new focus has stimulated intensive research into how specific patterns of entanglement can be created and manipulated in quantum many-body systems \cite{amico2008entanglement}.

One such controllable entanglement property is the scaling of the entanglement entropy, $S_{A}$, within a bipartite system. The ground states of local quantum lattice Hamiltonians typically obey an ‘area law' such that the entanglement entropy is proportional to the size of the boundary of the chosen subsystem, $A$ \cite{eisert2010colloquium, hastings2007area}. In 2010, Vitagliano, Riera and Latorre showed how tuning the coupling profile of the inhomogeneous XX model allows its ground state to transition smoothly from obeying an area law of entanglement entropy scaling to a volume law \cite{vitagliano2010volume}.
The ground state of this model is termed the ‘concentric singlet phase’ \cite{vitagliano2010volume} or simply ‘rainbow state’ \cite{ramirez2014conformal}, due to its distinctive structure of maximally entangled valence bonds connecting pairs of sites distributed symmetrically across the centre of the chain. This simple model hosts a rich variety of properties \cite{ramirez2020breaking, rodriguez2017more} and has been the subject of much interest in recent years \cite{de2020piercing, langlett2022rainbow, ramirez2015entanglement, pocklington2022stabilizing}. 

In this work we present a generalisation of the rainbow state model, whereby, with the introduction of staggered transverse field terms to the inhomogeneous XX model, the degree of entanglement between each concentric pair on the chain can be independently varied. Using a Real-Space Renormalization Group approach we derive recursive expressions for the induced effective coupling and transverse field terms. These expressions have an elegant description in terms of the formalism of $q$-deformed algebra \cite{manin1988quantum, fannes1996quantum}. For a chain of $2N$ sites the ground state is a tensor product of $N$ concentric $q$-deformed singlets, each with an associated deformation parameter, $q_{i}$, dependent on the transverse field and coupling parameters of our model. The variation of these physical parameters allows for the generation of any arbitrary set of single-particle entanglement energies.
To verify the validity of our results we perform a detailed numerical analysis. This analysis reveals that appropriate choices of the values of the transverse field parameter, and ordering of the degree of entanglement ensures a high fidelity between the exact ground state and the $q$-deformed rainbow. Moreover, we consider two special cases to demonstrate the applicability of our method. First, we consider the case $q_{1}=q_{2}=\dots =q_{N}$ such that each concentric pair has the same degree of entanglement. Second, we consider the case where the single-particle entanglement energies follow the ‘prime number spectrum’. This prime number decomposition employs the Moebius function to naturally mirror the entanglement spectra of free fermionic systems.

While our quantum simulator gives rise to effective couplings between sites on opposite ends of our chain, our model is completely local, given in terms of XX interactions and local magnetic fields. Thus, it directly lends itself to experimental verification and practical applications. Indeed, recent developments in cold atom experiments \cite{zeiher2015microscopic, parsons2015site, kennedy2015observation, baier2016extended, cheuk2015quantum, jotzu2014experimental} have offered unique opportunities to simulate such systems and access quantities related to entanglement \cite{abanin2012measuring, hauke2016measuring, pichler2016measurement, islam2015measuring, cardy2011measuring, alba2017out}. We expect that our quantum simulator can have direct applications in condensed matter or quantum technologies where specific structures of correlation patterns are requested between two subsystems.

\section{The $q$-Deformed Model}
In order to introduce our model for a chain of $2N$ spin-${1\over 2}$ particles, we first present a two-site Hamiltonian that allows for direct continuous variation of the degree of entanglement between its spins.

\subsection{Two Spins Hamiltonian}
\label{sec:2site}

\begin{figure}
  \includegraphics[width=\linewidth]{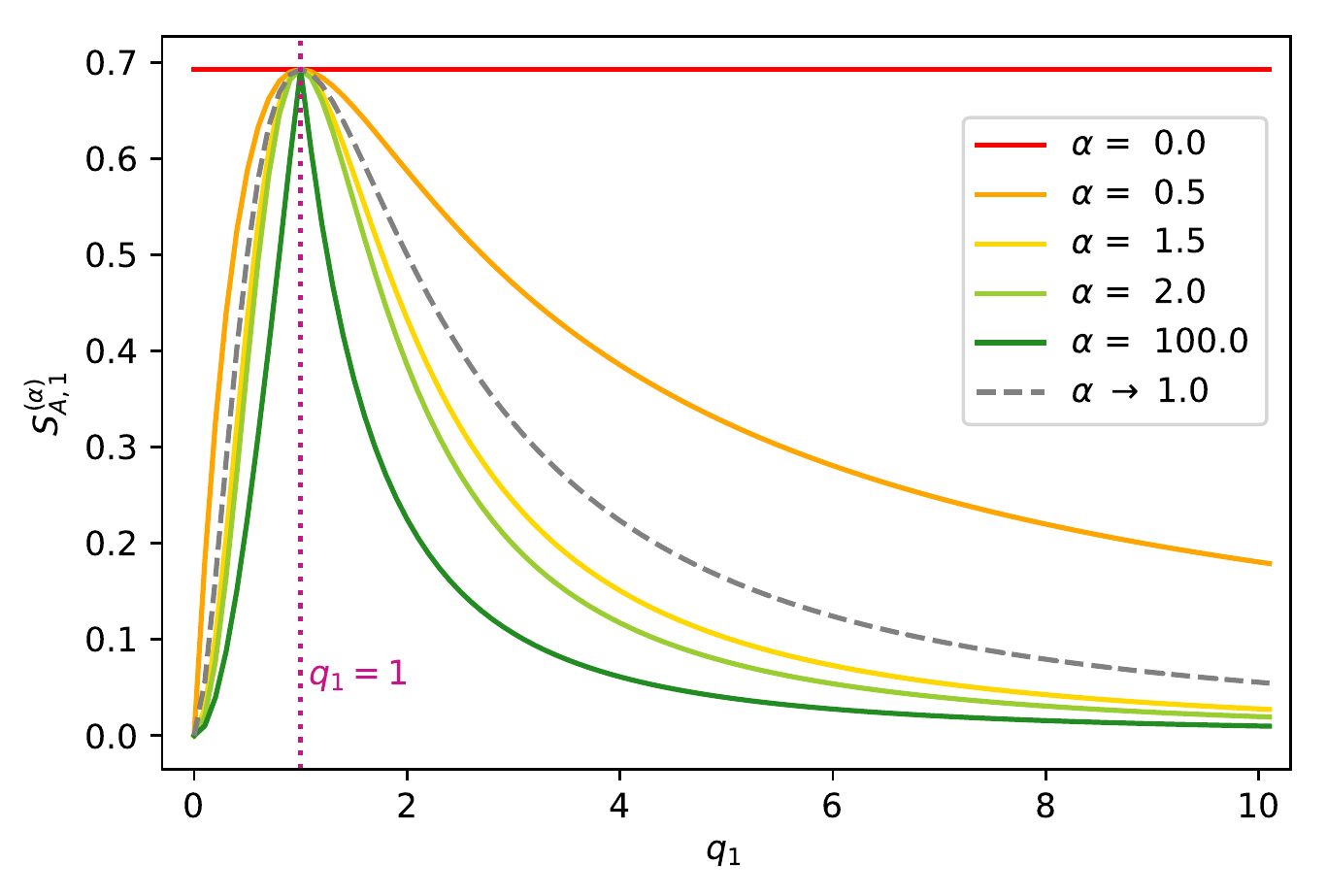}
\caption{The variation of the Renyi entropy, $S_{A,1}^{(\alpha)}$ for states given in \eqref{eq:SU(2)_q singlet} as a function of their deformation parameter $q_{1}$, for a range of fixed values of $\alpha$. The Renyi entropy takes maximal value $S_{A,1}^{(\alpha)}=\ln{2}$ when the deformation parameter $q_{1}=1,$ for all $\alpha$, and is unchanged under the transformation, $h_{1}\rightarrow -h_{1}$, such that $q_{1}\rightarrow\frac{1}{q_{1}}$, reflecting a symmetry of the Hamiltonian \eqref{eq:2-body}.
}
\label{fig:renyi}
\end{figure}

Consider the two-spins Hamiltonian
\begin{equation}
    \mathcal{H} = J_{1} \left(\sigma_{-1}^{x}\sigma_{1}^{x}+\sigma_{-1}^{y}\sigma_{1}^{y} \right) + h_{1} \left(\sigma_{-1}^{z} - \sigma_{1}^{z} \right).
    \label{eq:2-body}
\end{equation}
The ground state of $\mathcal{H}$ is given by
\begin{equation}
    \ket{\psi_{1}} = \frac{1}{\sqrt{[2]_{q_{1}}}}\left(q_{1}^{-\frac{1}{2}}\ket{\uparrow\downarrow}_{-1,1}-q_{1}^{\frac{1}{2}}\ket{\downarrow\uparrow}_{-1,1}\right),
    \label{eq:SU(2)_q singlet}
\end{equation}
and has ground state energy
\begin{equation}
	E_{1}=-[2]_{q_{1}}J_{1},
\end{equation}
where
\begin{equation}
    q_{1}=e^{\gamma_{1}}, \hspace{0.2cm} \sinh{\gamma_{1}}=\frac{h_{1}}{J_{1}},
    \label{eq:q1}
\end{equation}
and $[x]_{q}$ is the so-called quantum dimension
\begin{equation}
    [x]_{q} = \frac{q^{x}-q^{-x}}{q-q^{-1}}.
\end{equation}
This ground state is the singlet of the quantum group $SU(2)_{q_{1}}$ \cite{gomez1996quantum}. Such $q$-deformed valence bonds have been considered in relation to a range of quantum many-body models \cite{batchelor1994q, klumper1991equivalence, santos2012entanglement, batchelor1990q}, including the anisotropic $q$-deformed generalization of the spin-$1$ AKLT chain as considered in \cite{klumper1992groundstate, quella2020symmetry}.
In the limit $h_{1}\rightarrow 0$ such that $q_{1}\rightarrow 1$, we recover the maximally entangled singlet state of the standard $SU(2)$ Lie algebra. The degree of entanglement between this pair is directly related to the value of the deformation parameter, $q_{1}$, which is in turn directly related to our coupling and transverse field parameters via equation \eqref{eq:q1}. To investigate this, we bipartition the system down the centre of the chain into region $A$, and its complement $B$. The reduced density matrix of \eqref{eq:SU(2)_q singlet} is then determined for region $A$. The corresponding Renyi entropy of order $\alpha$ is given by
\begin{equation}
    S_{A,1}^{(\alpha)} = \frac{1}{1-\alpha} \ln{ \frac{1+q_{1}^{2\alpha}}{(1+q_{1}^{2})^{\alpha}}}, \hspace{0.5cm} \alpha>0, \alpha\neq 1
	\label{eq:Sq1}
\end{equation} 
and takes the maximum value $\ln{2}$ when $q_{1}=1$ for all $\alpha$ as shown in Figure \ref{fig:renyi}. This expression for the Renyi entropy reflects a symmetry of our model as under the transformation $h_{1}\rightarrow -h_{1}$, such that $q_{1}\rightarrow\frac{1}{q_{1}}$, the value of the Renyi entropy of order $\alpha$ is unchanged.

By considering the limit of $S_{A,1}^{(\alpha)}$ as $\alpha\rightarrow 1$ we obtain the expression for the von Neumann entanglement entropy of the pair
\begin{equation}
	S_{A,1} = \ln{(1+q_{1}^{2})}-\frac{q_{1}^{2}}{1+q_{1}^{2}} \ln{q_{1}^{2}}.
\end{equation}
This entropy can be varied continuously to achieve all values in the maximal range $0\leq S_{A,1}\leq \ln{2}$, by varying $0\leq q_{1}\leq \infty$, or equivalently $-\infty\leq \frac{h_{1}}{J_{1}}\leq \infty$. We see that by varying the physical parameters of our model we can achieve all degrees of pairwise entanglement between the two spins. 

\subsection{$2N$ Spin Hamiltonian}

\begin{figure}
  \includegraphics[width=\linewidth]{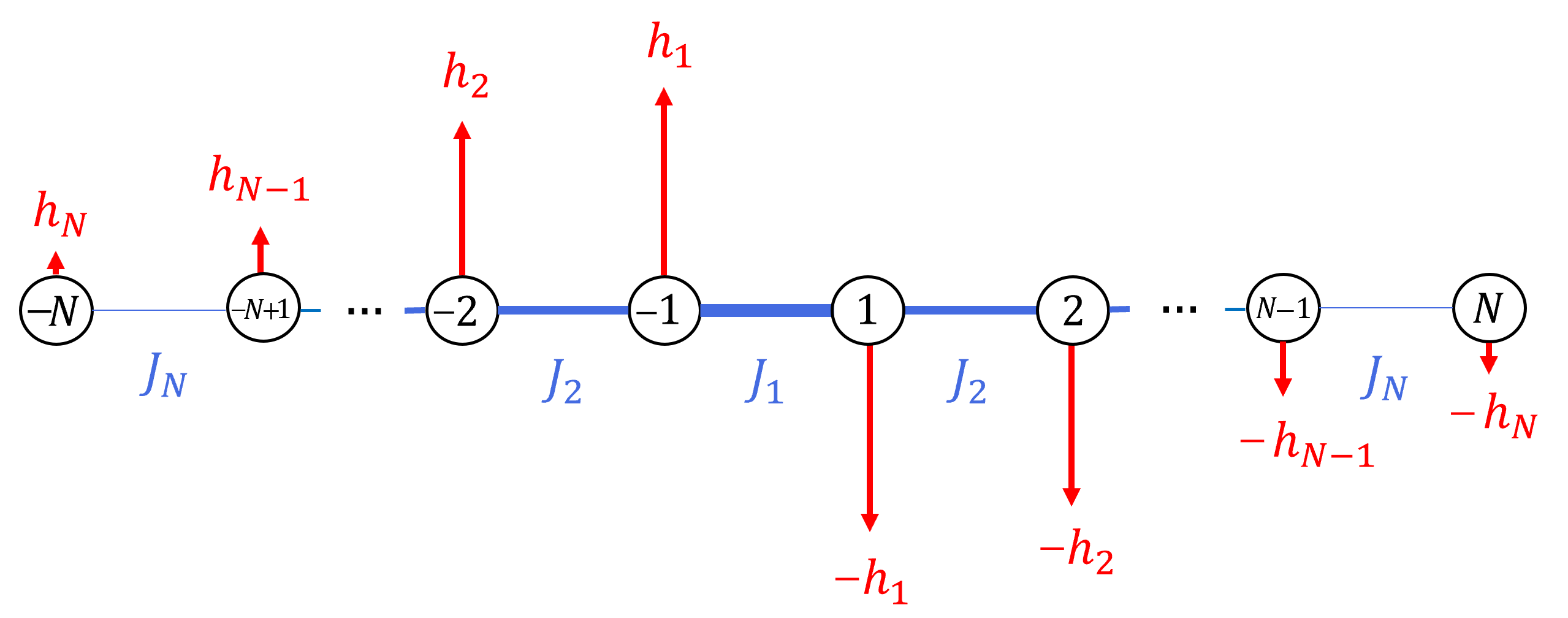}
\caption{The $q$-deformed spin model for a chain of $2N$ sites. The blue lines represent the XX coupling terms, $J_{i}$, and the red arrows represent the magnitude and direction of the transverse magnetic field, $h_{i}$. 
The magnitude of both the coupling and transverse field are symmetric about the centre of the chain, with decreasing strength moving outwards.
}
\label{fig:labelling}
\end{figure}

The simple two-spin Hamiltonian presented above is the basis on which we construct our general model for a chain of any even number of spins.
We now consider a chain of $2N$ spin-$\frac{1}{2}$ particles with the following Hamiltonian
\begin{multline}
     \mathcal{H} = \sum_{i=1}^{N} h_{i} \left(\sigma_{-i}^{z} - \sigma_{i}^{z} \right) + J_{1} \left(\sigma_{-1}^{x}\sigma_{1}^{x}+\sigma_{-1}^{y}\sigma_{1}^{y} \right)  \\ + 
     \sum_{i=2}^{N} J_{i}\left(\sigma_{-i}^{x}\sigma_{-(i-1)}^{x} + \sigma_{-i}^{y}\sigma_{-(i-1)}^{y} + \sigma_{i-1}^{x}\sigma_{i}^{x} + \sigma_{i-1}^{y}\sigma_{i}^{y} \right).
     \label{eq:model}
\end{multline}
We have introduced the site labelling $\{ -N, -(N-1), \dots, -2, -1, 1, 2, \dots, N-1, N\}$ such that sites $-i$ and $i$ are equidistant from a central bipartition of the chain, as shown in Figure \ref{fig:labelling}.
In order to find the ground state of our model we have used the Real-Space Renormalization Group approach as first introduced by Ma and Dasgupta in \cite{dasgupta1980low} and later developed by Fisher with the application of the method to the Random Transverse Field Ising Chain \cite{fisher1995critical}. This approach allows us to consider the ground state properties of random quantum chains by iteratively decimating the degrees of freedom with highest energy in order to derive an overall effective low-energy model. We start by first briefly reviewing this method with reference to the known results of the $h_{i}\rightarrow0$ limit of our model.



\subsubsection{XX Model Renormalization Group}

In the limit, $h_{i}=0$, our Hamiltonian \eqref{eq:model} is equivalent to that of the inhomogeneous XX model acting on a chain of $2N$ spins
 \begin{equation}
     H_{XX} = \sum_{i=1}^{2N} J_{i} \left(\sigma_{i}^{x}\sigma_{i+1}^{x}+\sigma_{i}^{y}\sigma_{i+1}^{y} \right),
 \end{equation}
where we have re-adopted the standard site labelling $\{1, 2, \dots, 2N-1, 2N\}$. Using the Renormalization Group (RG) approach for some random coupling profile, the highest energy term such that $J_{i} \gg J_{i-1}, J_{i+1}$, is identified and diagonalised independently of the rest of the chain. To zeroth-order in perturbation theory, the ground state of the system is then
 \begin{equation}
     \ket{\psi} = \ket{\psi_{j<i}}\otimes\ket{\psi^{-}_{i}}\otimes\ket{\psi_{j>i}},
 \end{equation}
 where $\ket{\psi^{-}_{i}} = \frac{1}{\sqrt{2}}(\ket{\uparrow\downarrow}_{i,i+1}-\ket{\downarrow\uparrow}_{i,i+1})$ is the maximally entangled singlet ground state of the two-site XX model and $\ket{\psi_{j<i}}$, $\ket{\psi_{j>i}}$ refer to the state of the spins to the left and right of the singlet, respectively. To compute higher order corrections to the ground state of our system we initial consider the spins $i$ and $i+1$ to be ‘frozen’ into this singlet state. Then we employ perturbation theory to find the effect induced by quantum fluctuations on the neighbouring spins, as shown in \cite{vitagliano2010volume}. It is found that an effective coupling arises between sites $i-1$ and $i+2$ of strength
 \begin{equation}
     \Tilde{J}_{i-1,i+2} = \frac{J_{i-1}J_{i+1}}{J_{i}}.
	\label{eq:XXrescaling}
 \end{equation}
 In this way the coupling between sites $i$ and $i+1$ is replaced by effective longer range interaction that captures the low-energy properties of the model. For a random coupling profile, successive iterations of this procedure yield a ‘random singlet phase', as singlets form between the pairs of spins most strongly coupled after each decimation. In \cite{vitagliano2010volume} Vitagliano, Riera and Latorre demonstrated how a coupling profile that decays exponentially away from the centre of the chain produces a special form of ground state known as the ‘concentric singlet phase'. This ground state is also known as the ‘rainbow state', due to it's distinctive structure of a series of singlets symmetrically distributed around the centre of the chain. 
For any given bipartition, the entanglement entropy is directly proportional to the number of singlets ‘cut' by the bipartition. Thus, for such a coupling profile, the area law of entanglement entropy is maximally violated.



\subsubsection{$q$-Deformed Model Renormalization Group}

\begin{figure}
  \includegraphics[width=0.6\linewidth]{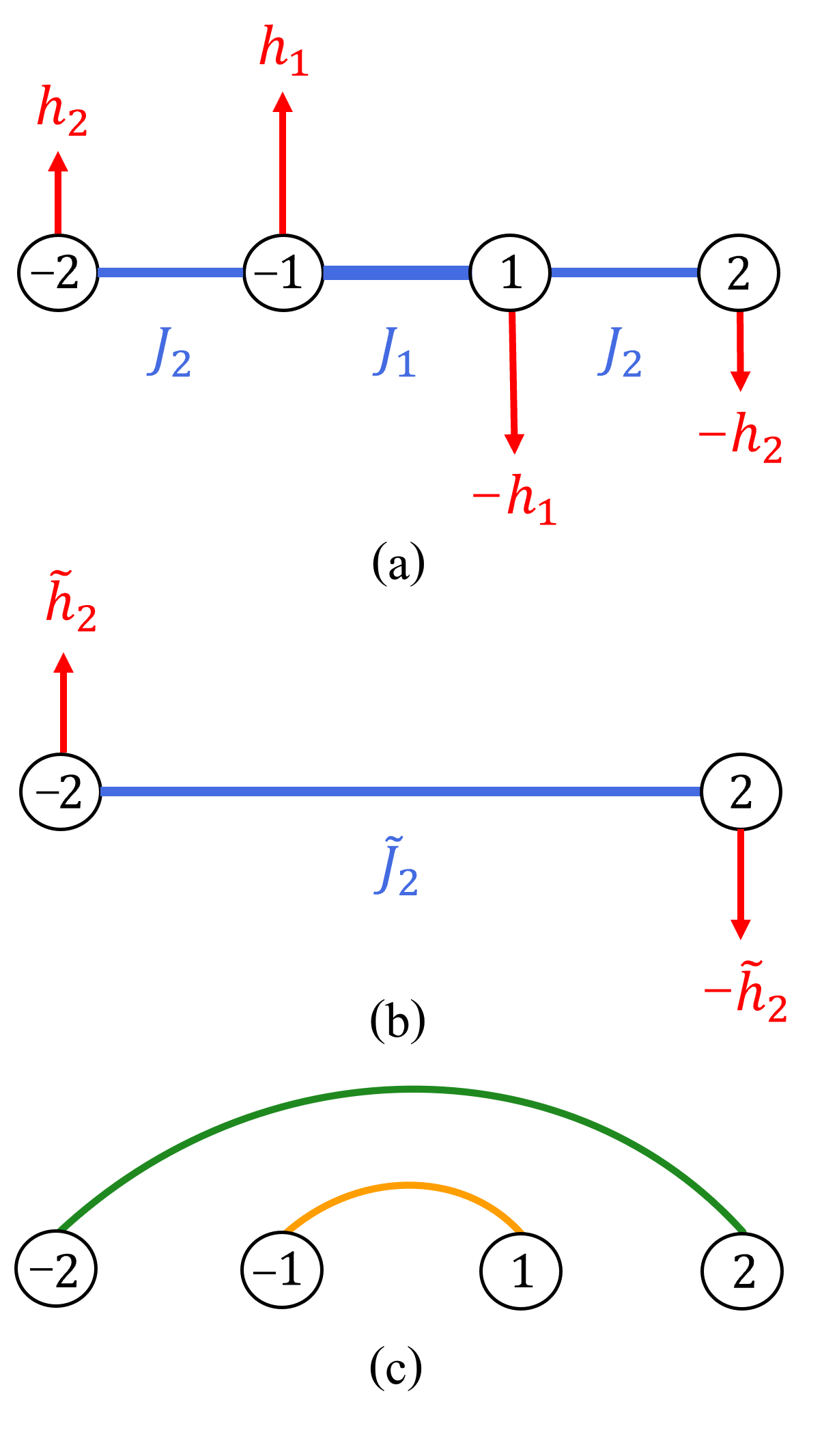}
\caption{The Real-Space RG procedure. (a) Our model \eqref{eq:model} acting on a chain of four spins. For $J_{1},h_{1}\gg J_{2},h_{2}$ perturbation theory yields a $q$-deformed singlet \eqref{eq:q1} between the central two spins. (b) These spins are integrated out and an effective Hamiltonian of the form \eqref{eq:2-body} is found to act between sites $-2$ and $2$ with renormalized coupling $\Tilde{J_{2}}$ and transverse field $\Tilde{h}_{2}$. (c) Diagonalization of this effective Hamiltonian yields the ground state $\ket{\psi}=\ket{\psi_{1}}\otimes\ket{\psi_{2}}$. The difference in colour of the bonds between the two pairs indicates the difference in correlations that can be achieved by appropriately tuning $J_{1}, J_{2}, h_{1}$ and $h_{2}$.
}
\label{fig:4RG}
\end{figure}

We now apply the Real-Space RG approach to the generalised model defined in \eqref{eq:model}. In the limit $J_{1}, h_{1} \gg J_{2}, h_{2}$, this yields the ground state
\begin{equation}
    \ket{\psi} = \ket{\psi_{i<-1}}\otimes\ket{\psi_{1}}\otimes\ket{\psi_{i>1}},
\end{equation}
to zeroth-order in perturbation theory, where $\ket{\psi_{1}}$ is the $q_{1}$-deformed singlet as defined in equation \eqref{eq:q1}. To compute corrections to the ground state at higher orders, second-order perturbation theory is used, as illustrated in Figure \ref{fig:4RG} (see also Appendix \ref{sec:4sitePT}). We derive an effective Hamiltonian of the form \eqref{eq:2-body} acting between sites $-2$ and $2$ with a renormalized coupling 
\begin{equation}
	\Tilde{J_{2}} = \frac{4J_{2}^{2}}{[2]_{q_{1}}^{2}J_{1}},
		\label{eq:J2}
\end{equation}
and transverse field terms
\begin{equation}
	\Tilde{h}_{2} = h_{2} - \frac{2\left(q_{1}-\frac{1}{q_{1}}\right)J_{2}^{2}}{[2]_{q_{1}}^{2}J_{1}}.
	\label{eq:h2}
\end{equation}
In the case that $\Tilde{J_{2}}, \Tilde{h}_{2} \gg J_{3}, h_{3}$ this effective Hamiltonian can be diagonalised to yield an additional $q_{2}$-deformed singlet, $\ket{\psi_{2}}$, between sites $-2$ and $2$, where $q_{2}=e^{\gamma_{2}}, $ $\sinh{\gamma_{2}}=\frac{\Tilde{h}_{2}}{\Tilde{J_{2}}}$. 

If the couplings throughout the chain are selected such that $\Tilde{J_{i}}, \Tilde{h}_{i} \gg J_{i+1}, h_{i+1}$, then repeated iterations of this renormalization process will eventually yield the overall ground state
\begin{equation}
    \ket{\psi} = \ket{\psi_{1}}\otimes\ket{\psi_{2}}\otimes\dots\otimes\ket{\psi_{N}},
    \label{eq:GS}
\end{equation}
where
\begin{equation}
    \ket{\psi_{i}} = \frac{1}{\sqrt{[2]_{q_{i}}}}\left(q_{i}^{-\frac{1}{2}}\ket{\uparrow\downarrow}_{-i,i}-q_{i}^{\frac{1}{2}}\ket{\downarrow\uparrow}_{-i,i}\right),
    \label{eq:SU(2)_qn}
\end{equation}
and for $i>1$
\begin{equation}
    q_{i}=e^{\gamma_{i}} \text{ with } \sinh{\gamma_{i}}=\frac{\Tilde{h}_{i}}{\Tilde{J_{i}}}.
    \label{eq:qi}
\end{equation}
The renormalized coupling and transverse field parameters for the effective Hamiltonian between spins $-i$ and $i$ are given by the recursive expressions
\begin{equation}
 \Tilde{J_{i}} = \frac{4J_{i}^{2}}{[2]_{q_{i-1}}^{2}\Tilde{J}_{i-1}},
    \label{eq:ji'}
\end{equation}
\begin{equation}
    \Tilde{h}_{i} = h_{i} - \frac{2\left(q_{i-1}-\frac{1}{q_{i-1}}\right)J_{i}^{2}}{[2]_{q_{i-1}}^{2}\Tilde{J}_{i-1}}.
    \label{eq:hi'}
\end{equation}
From equations \eqref{eq:ji'} and \eqref{eq:hi'}, we see that the expressions for $\Tilde{J_{i}}$ and $\Tilde{h}_{i}$ are dependent on all previous $\Tilde{J}_{j<i}$, $\Tilde{h}_{j<i}$. By fixing all previous $i-1$ values, it is always possible to vary the associated physical parameters $J_{i}$ and $h_{i}$ such as to achieve any $1\leq q_{i}\leq \infty$. In this way, we will show that the deformation of each $q$-singlet can be individually tuned to achieve any degree of pairwise entanglement between a given pair of spins.

\section{Entanglement Properties of the $q$-Deformed Rainbow}

The ground state \eqref{eq:GS} of our Hamiltonian \eqref{eq:model} has a tensor product form. Subsequently, the reduced density matrix across a central bipartition admits the tensor product decomposition 
\begin{equation}
    \rho_{A} = \rho_{1}\otimes \rho_{2} \otimes \dots \otimes \rho_{N},
\end{equation}
where each $\rho_{i}$ is diagonal, given by
\begin{equation}
    \rho_{i} = \begin{pmatrix}
    \frac{1}{1+q_{i}^{2}} & 0 \\
    0 & \frac{q_{i}^{2}}{1+q_{i}^{2}}
    \end{pmatrix}.
\end{equation}
This decomposition yields simple expressions for many of the entanglement properties of the $q$-deformed rainbow, as we will see in the following.

\subsection{Renyi and von Neumann entropies}
\label{sec:entropy}

Using the reduced density matrix tensor product decomposition we derive the form of the Renyi entropy of order $\alpha$ of the ground state \eqref{eq:GS} across a central bipartition
\begin{equation}
    S_{A}^{(\alpha)} = \frac{1}{1-\alpha} \sum_{i=1}^{N}\ln{ \frac{1+q_{i}^{2\alpha}}{(1+q_{i}^{2})^{\alpha}}}, \hspace{0.3cm} \alpha>0,  \alpha\neq 1.
\end{equation}
In the limit $\alpha \rightarrow 1$ we obtain an expression for the von Neumann entropy of the ground state
\begin{equation}
	S_{A} = -\sum_{i=1}^{N} \left[ \ln{(1+q_{i}^{2})}-\frac{q_{i}^{2}}{1+q_{i}^{2}} \ln{q_{i}^{2}} \right].
    \label{eq:totalEE}
\end{equation}
In Section \ref{sec:2site} we found the von Neumann entropy, $S_{A,1}$, of a single pair of spins as a function of the deformation parameter $q_{1}$, as given by \eqref{eq:Sq1}. By extending this definition to that of the von Neumann entropy of the state $\ket{\psi_{i}}$ between spins $-i$ and $i$
\begin{equation}
	S_{A,i} = \ln{(1+q_{i}^{2})}-\frac{q_{i}^{2}}{1+q_{i}^{2}} \ln{q_{i}^{2}},
	\label{eq:indEE}
\end{equation}
it is clear that the total von Neumann entropy is a sum of the individual von Neumann entropies of each concentric pair of spins on the chain. This is also true for the Renyi entropy, and is a natural consequence of the tensor product form of the reduced density matrix.  By independently varying each $q_{i}$, we can therefore achieve all degrees of entanglement in the allowed maximal range $0\leq S_{A}\leq N\ln{2}$.

\subsection{Entanglement Spectrum}
\label{sec:entspec}

The entanglement spectrum was introduced by Li and Haldane \cite{li2008entanglement} as an alternative entanglement measure that aimed to capture a complete representation of the entanglement between two subsystems \cite{calabrese2008entanglement}. The values of the spectrum, $E_{i}$, are related to the eigenvalues of the reduced density matrix, $\lambda_{i}$, via
\begin{equation}
    \lambda_{i} = e^{-E_{i}}.
    \label{eq:ES}
\end{equation}
The exponential relationship means that the dominant quantum correlations depend predominantly on the ‘lowest' part of the entanglement spectrum.

The entanglement spectrum reflects many of the physical properties of the system \cite{alba2012boundary, metlitski2011entanglement, yang2015two, leiman2015correspondence, geraedts2016many} and serves as a fingerprint of topological order \cite{thomale2010entanglement, chandran2011bulk, qi2012general}. For any non-interacting model, Wick's theorem shows that the spectrum can be constructed from a set of single-particle entanglement energies as
\begin{equation}
    E^{f}_{j}(\epsilon) = E_{0} + \sum_{i=1}^{N} n_{i}(j) \epsilon_{i},
    \label{eq:free}
\end{equation}
where $E_{0}$ is a normalization constant and each $n_{j}=\{0,1\}$ \cite{peschel2003calculation}.

For our $q$-deformed rainbow, we find that
\begin{equation}
    E_{0} = \sum_{i=1}^{N} \ln{(1+q_{i}^{2})},
\end{equation}
and
\begin{equation}
    \epsilon_{i} = -\ln{q_{i}^{2}}.
	\label{eq:epsilon_i}
\end{equation}
Hence, the deformation parameters, $q_{i}$, of the $q$-deformed singlets directly determine the single-particle entanglement energies. As each $q_{i}$ can take any value in the range $0\leq q_{i}\leq\infty$, each $\epsilon_{i}$ can be individually tuned to take any value $-\infty\leq\epsilon_{i}\leq\infty$. 
\begin{figure}
  \includegraphics[width=\linewidth]{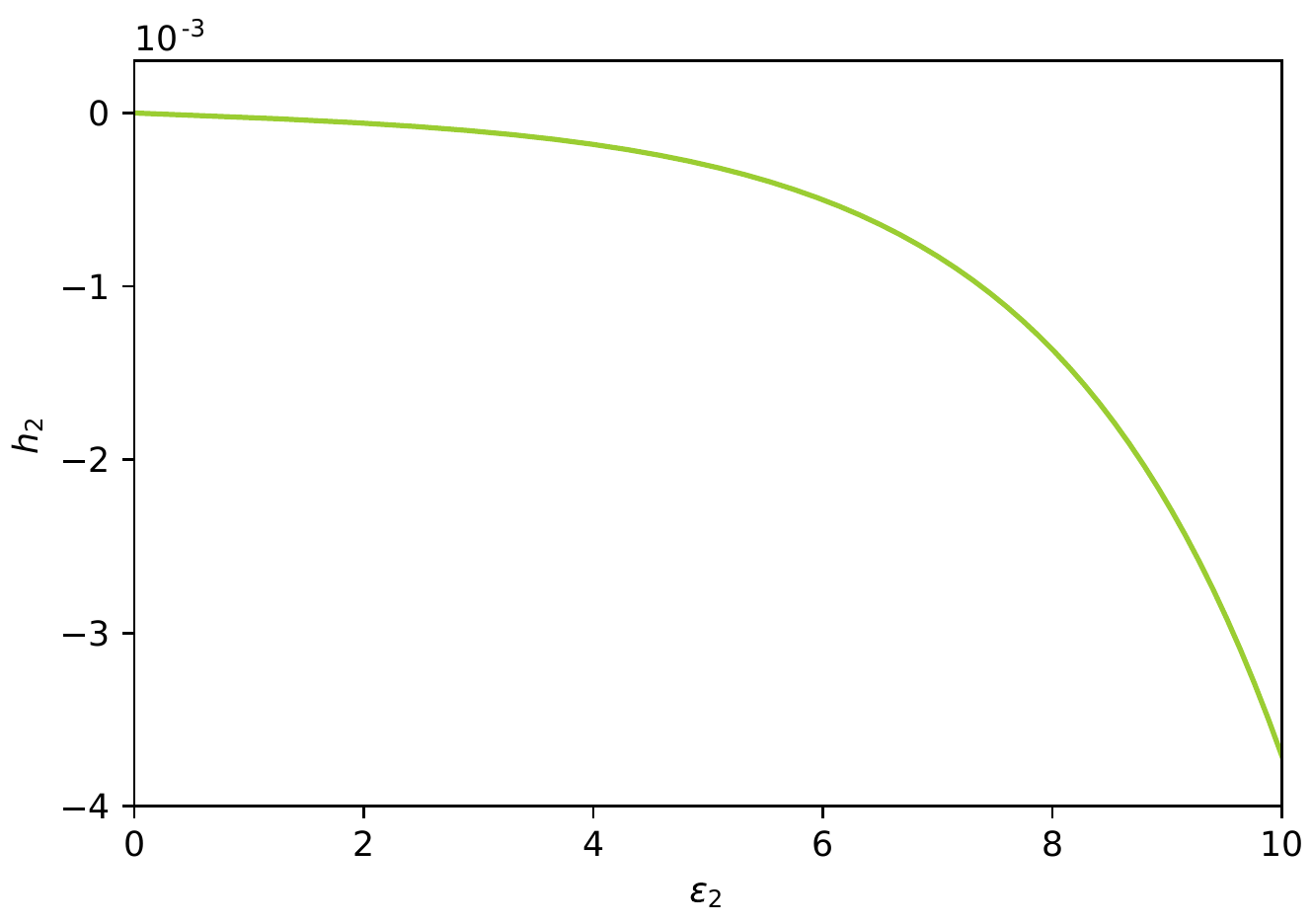}
\caption{The values of $h_{2}$ required to generate any $0\leq\epsilon_{2}\leq10$ for fixed $J_{1}=h_{1}=1$, $J_{2}=0.01$ ($\epsilon_{1}=-2\sinh^{-1}{(1)}$). Any desired value of $\epsilon_{2}$ in this range can be obtained by selecting the corresponding value of $h_{2}$.
}
\label{fig:h2fore2}
\end{figure}

By combining \eqref{eq:qi} and \eqref{eq:epsilon_i} we derive the simple relationship
\begin{equation}
	\gamma_{i}=-\frac{\epsilon_{i}}{2}.
\end{equation}
This in turn yields an expression for the required ratio of the renormalized parameters for a given pair in order to produce a specific desired single-particle entanglement energy
\begin{equation}
\frac{\Tilde{h}_{i}}{\Tilde{J_{i}}} = -\sinh{\left(\frac{\epsilon_{i}}{2}\right)}.
\label{eq:singpart}
\end{equation}
As a result, each single particle energy of the entanglement spectrum can be directly obtained by appropriately tuning a single effective magnetic field. In Appendix \ref{sec:entenergies_app} we expand these expressions to derive closed forms for the required ratio of the physical coupling parameters. In Figure \ref{fig:h2fore2} the dependence of $\epsilon_{2}$ on $h_{2}$ for fixed $J_{1},h_{1}$ and $J_{2}$ is illustrated. For the shown range, any desired $\epsilon_{2}$ can be simulated by simply reading off the corresponding value of $h_{2}$. In this way, by fixing all previous $i-1$ single-particle entanglement energies, $\epsilon_{i}$ can be tuned to any desired value by appropriately varying $h_i$.

\section{Fidelity Optimisation}

In the previous Section, we have demonstrated how controlled variation of the parameters of our model in the strong inhomogeneity limit $\Tilde{J_{i}}, \Tilde{h}_{i}\gg J_{i+1}, h_{i+1}$ allows for the generation of any arbitrary pattern of correlations given in terms of the entanglement entropy \eqref{eq:totalEE} or the single-particle entanglement energies \eqref{eq:singpart}. In this Section we present how the parameters of our model can be chosen such that the fidelity is maximised for any desired entanglement profile. In quantum information theory, fidelity is a measure of the ‘closeness' of two quantum states, $\ket{\psi_{A}}$ and $\ket{\psi_{B}}$, given by the squared overlap, $\abs{\bra{\psi_{A}}\ket{\psi_{B}}}^{2}$ \cite{nielsen2002quantum2}. To optimise the choice of parameters for any desired correlation profile, we consider the variation of the fidelity between the exact ground state of our model and the $q$-deformed rainbow in the case $N=4$. 

\subsection{Optimising $h_{2}$}

\begin{figure}
  \includegraphics[width=\linewidth]{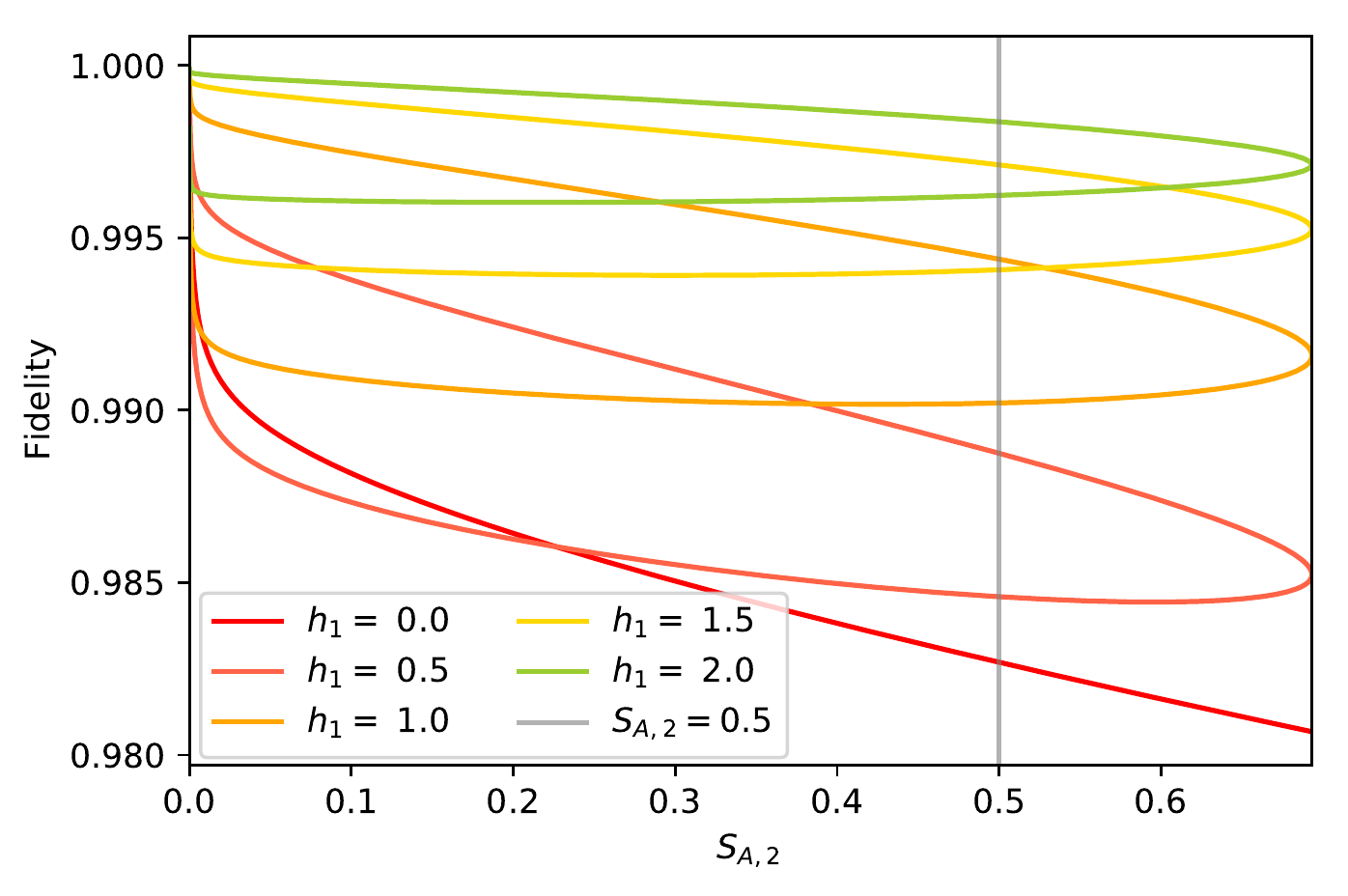}
\caption{Variation of the ground state fidelity with the entanglement entropy of the outer pair for different fixed values of $h_{1}$ with $J_{1}=1$ and $J_{2}=0.1$. The grey line shows an example of a desired outer entanglement entropy, $S_{A,2}=0.5$. The two intersections with each curve for $h_{1}>0$ indicate two possible values of $h_{2}$ to generate the desired $S_{A,2}$ with a distinct difference in fidelity. This choice can be used to optimise the accuracy of our model.
}
\label{fig:varyh2}
\end{figure}

In Section \ref{sec:2site}, we noted that a symmetry of our two-site Hamiltonian \eqref{eq:2-body} results in the preservation of the von Neumann entropy, $S_{A,1}$, under the transformation $h_{1}\rightarrow -h_{1}$.
Here, we will show that although $S_{A,2}$ possesses a similar symmetry under the transformation $\Tilde{h}_{2}\rightarrow-\Tilde{h}_{2}$, one of these values will yield a significantly higher fidelity than the other corresponding to the choice of sign of $\frac{h_{1}}{J_{1}}$. 

\begin{figure}
  \includegraphics[width=\linewidth]{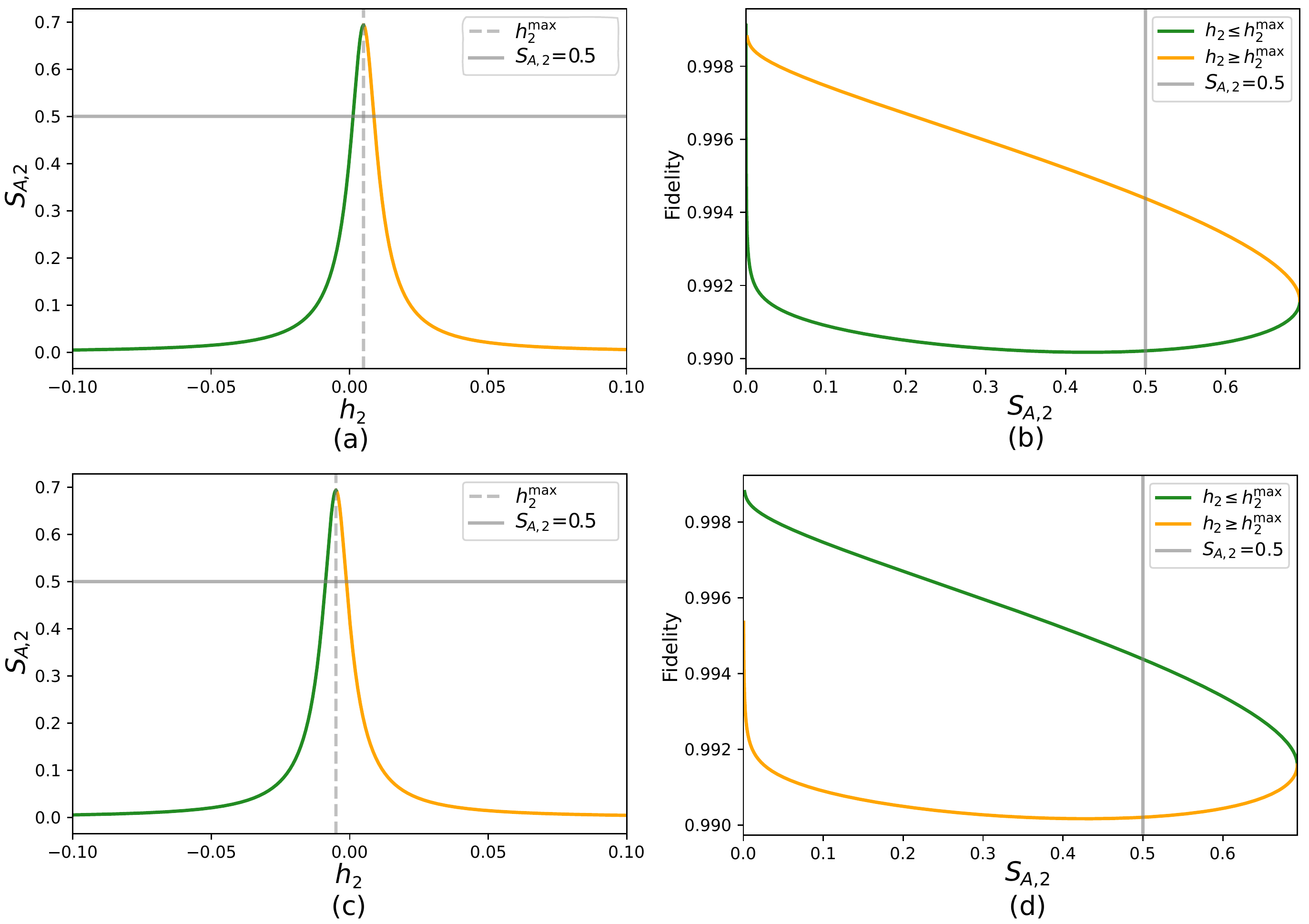}
\caption{(a) Variation of the entanglement entropy of the outer pair with $h_{2}$ for $J_{1}=h_{1}=1$ and $J_{2}=0.1$. (b) The corresponding variation of the ground state fidelity with the entanglement entropy for both the values less than and greater than $h_{2}^\text{max}$. The higher fidelity branch corresponds to the values $h_{2}\geq h_{2}^\text{max}$. (c) Variation of the entanglement entropy of the outer pair with $h_{2}$, now for $J_{1}=1$, $J_{2}=0.1$ and $h_{1}=-1$. (d) The corresponding variation of the ground state fidelity with the entanglement entropy for both the values less than and greater than $h_{2}^\text{max}$. The higher fidelity branch now corresponds to the values $h_{2}\leq h_{2}^\text{max}$. On all four subfigures the solid grey line serves to illustrate this fidelity optimisation for the specific case $S_{A,2}=0.5$.
}
\label{fig:4figs}
\end{figure}

In Figure \ref{fig:varyh2} we plot the fidelity between the $q$-deformed rainbow and the exact ground state of \eqref{eq:model} as a function of the entanglement entropy between sites $-2$ and $2$ for a range of constant values of $h_{1}$. For each curve $J_{1}, h_{1}$ and $J_{2}$ are fixed such that $S_{A,2}$ is a function of $h_{2}$. We see that for any desired value of $S_{A,2}$, for example $S_{A,2}=0.5$ as indicated by the vertical grey line, the two intersections with each curve indicate two values of $h_{2}$ that correspond to the same degree of entanglement, but with a distinct difference in fidelity. As described, these two solutions arise due to the natural symmetry of the entanglement entropy about the value of $h_{2}$ yielding maximal entanglement between sites $-2$ and $2$. To find the value $h_2^\text{max}$ that maximises $S_{A,2}$ we set $\Tilde{h}_{2}=0$ in equation \eqref{eq:h2} and obtain
\begin{equation}
	h_{2}^\text{max} = \frac{2\left(q_{1}-\frac{1}{q_{1}}\right)J_{2}^{2}}{[2]_{q_{1}}^{2}J_{1}}.
\end{equation}
The symmetry of the entanglement entropy $S_{A,2}$ about $h_{2}^\text{max}$ is shown in Figure \ref{fig:4figs}(a) for the case $J_{1}=h_{1}=1$, $J_{2}=0.1$. By mapping these values onto the plot of fidelity with $S_{A,2}$ as shown in Figure \ref{fig:4figs}(b), we see that we have a ‘high fidelity branch’ corresponding to $h_{2}\geq h_{2}^\text{max}$ and a ‘low fidelity branch’ for $h_{2}\leq h_{2}^\text{max}$. For any desired value of $S_{A,2}$, the fidelity is clearly maximised by choosing the appropriate value of $h_{2} \geq h_{2}^\text{max}$. In contrast, if $\frac{h_{1}}{J_{1}}$ is negative as shown in Figure \ref{fig:4figs}(c) and (d), the opposite is true, and the fidelity is maximised by selecting the value of $h_{2}$ from the branch $h_{2}\leq h_{2}^\text{max}$. In this way, for fixed couplings $J_{1}$ and $J_{2}$, the direction of the magnetic fields applied to sites $i=-1,1$  dictate the magnitude and direction of the magnetic field that should be applied to sites $i=-2,2$ in order to maximise the accuracy of our model. 

\subsection{Optimising Order of Pairs}
\label{sec:ordering}

\begin{figure}
  \includegraphics[width=\linewidth]{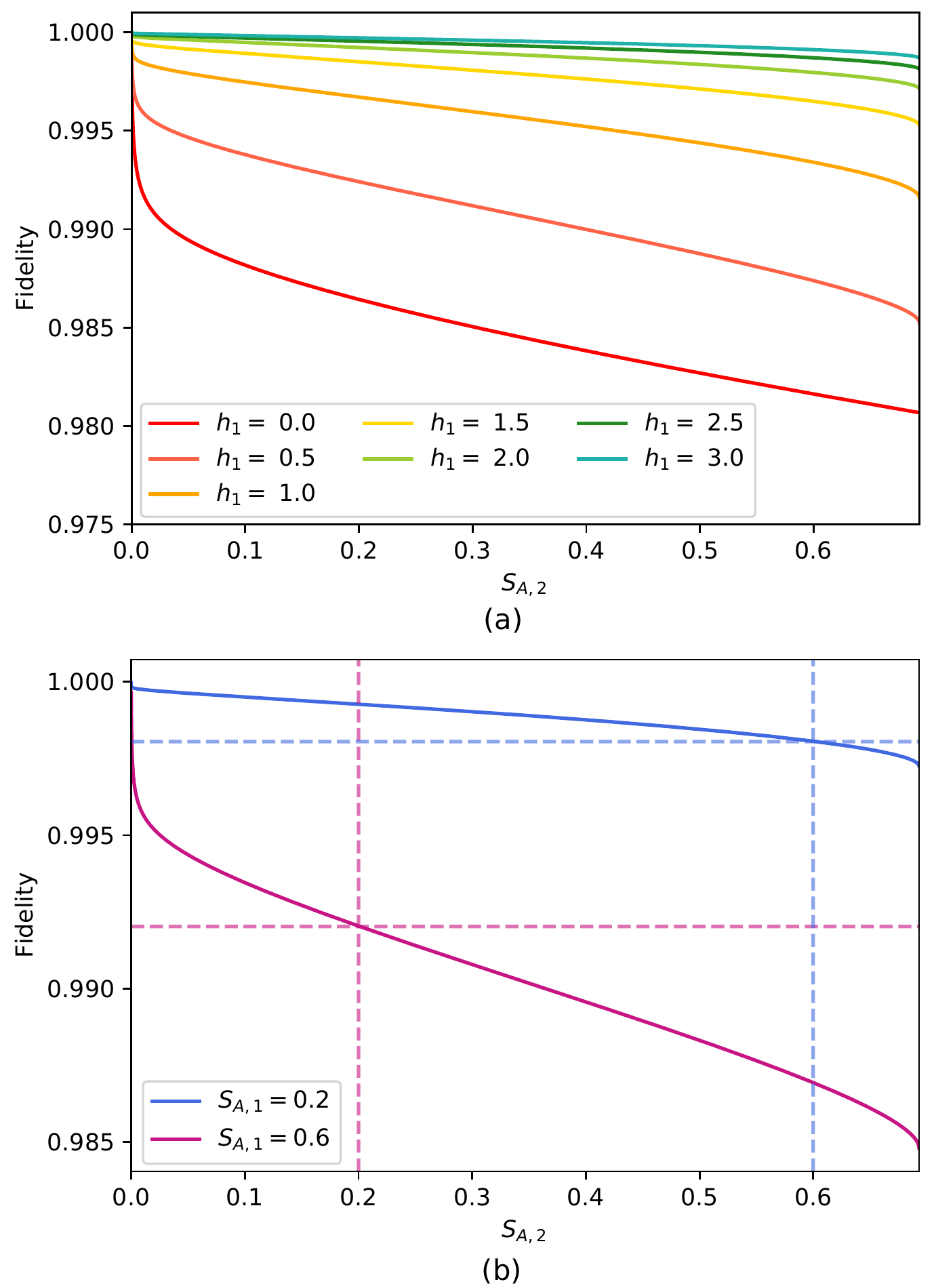}
\caption{(a) The variation of the fidelity with $S_{A,2}$ for a range of values of $h_{1}$ and fixed $J_{1}=1$, $J_{2}=0.1$. We observe that always the fidelity decreases as $S_{A,2}$ increases. For two curves with different value of $h_{1}$, it is observed that all values of $S_{A,2}$ can be produced with a higher fidelity by the curve with a higher value of $h_{1}$ or equivalently a lower value of $S_{A,1}$. The implications of this on producing some desired set of von Neumann entropies are examined further in (b). Here the dashed lines serve to illustrate the higher value of fidelity achieved by choosing $h_{1}$ and $h_{2}$ to obtain $S_{A,1}=0.2, S_{A,2}=0.6$ as opposed to $S_{A,1}=0.6, S_{A,2}=0.2$. In general, the fidelity is optimised by selecting the parameters of our model such that $S_{A,i}\leq S_{A,i+1}$.
}
\label{fig:fidlines}
\end{figure}

Our system has a symmetry with respect to which pair $i,-i$ of spins is used to tune a certain single particle entanglement energy $\epsilon_k$. We can use this freedom, in conjunction with the optimisation procedure of the previous Subsection to optimise the overall fidelity of our chain simulator. To proceed, we adopt the appropriately restricted range of $h_{2}\geq h_2^\text{max}$ that maximises the fidelity of two pairs of spins. We then consider the variation of the fidelity with $h_{2}$ for a range of fixed values of $h_{1}$ as shown in Figure \ref{fig:fidlines}(a). Here $\frac{h_{1}}{J_{1}}>0$ for each curve such that we have selected the values $h_{2}\geq h_{2}^\text{max}$. We first observe that for each fixed value of $h_{1}$, as $S_{A,2}$ increases, the fidelity decreases. This result is more notable for low values of $h_{1}$, such that the combination of parameters with lowest fidelity corresponds to the reproduction of the rainbow state with $h_{1}=h_{2}=0$. Hence, our model is best at accurately producing lower degrees of entanglement between the concentric pairs of sites.

\begin{figure}
  \includegraphics[width=\linewidth]{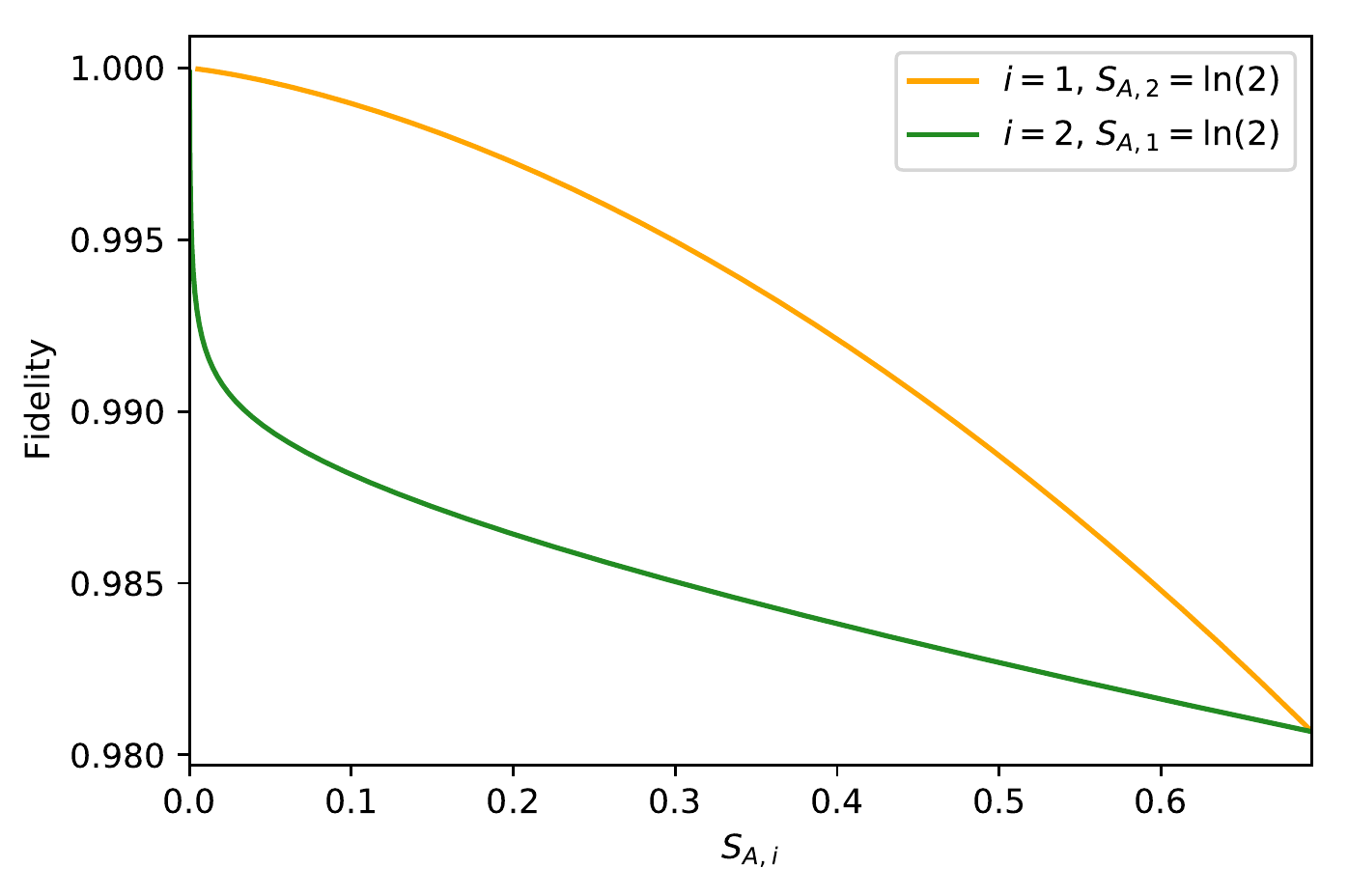}
\caption{The fidelity of the four-site $q$-deformed rainbow and the exact ground state of our model  when one pair is maximally entangled and the entanglement entropy of the other is varied ($J_{1}=1, J_{2}=0.1$).
For every combination of entanglement entropies, the fidelity is maximised by choosing our parameters such that the maximally entangled pair is between sites $-2$ and $2$.
}
\label{fig:fid_othermax}
\end{figure}

For fixed $J_{1}$, a larger magnitude of $h_{1}$ coresponds to a lower value of $S_{A,1}$. Figure \ref{fig:fidlines}(a) therefore also shows that as the value of $h_{1}$ increases and $S_{A,1}$ decreases, the fidelity with which any desired $S_{A,2}$ can be achieved increases. Consider the case in which we want to use our model to generate a given pair of two-site von Neumann enanglement entropies, for example, $S_{A,i}=0.2$ and $S_{A,j}=0.6$. Our simulator has the freedom in the choice $i=1, j=2$ or $i=2, j=1$. We will employ this freedom to choose the combination that maximises the fidelity. In Figure \ref{fig:fidlines}(b) we plot the two curves corresponding to $S_{A,1}=0.2$, and $S_{A,1}=0.6$. It is clear from this plot that the fidelity is always maximum by choosing the parameters such that $S_{A,1}=0.2, S_{A,2}=0.6$. In general it is always true that the fidelity is maximised by ordering the pairs such that $S_{A,i}\leq S_{A,i+1}$. Figure~\ref{fig:fid_othermax} further illustrates this with a direct comparison of the fidelity with the degree of entanglement between one pair when the other is maximally entangled. For all values of $S_{A,i}$, we observe that the fidelity is maximised when the maximally entangled state lies between sites $-2$ and $2$. 

By combining equations \eqref{eq:indEE} and \eqref{eq:epsilon_i}, we see that the condition $S_{A,i}\leq S_{A,i+1}$ is equivalent to $\abs{\epsilon_{i}}\geq \abs{\epsilon_{i+1}}$. For some desired set of single-particle entanglement energies, $\{\epsilon_{i}\}$, our model allows for complete freedom in assigning which pair of sites corresponds to a given energy. In implementing our model we therefore choose to tune the parameters 
such that the magnitude of the single-particle energy generated by sites $-i$ and $i$ decreases with increasing $i$ in order to optimise the fidelity.

\section{Special Cases}

We have shown how the tuning of the parameters of our model allows for the generation of any arbitrary set of single-particle entanglement energies. In this section we highlight two interesting applications: the case in which all deformation parameters are equal, and the reproduction of the single-particle entanglement energies for the ‘prime number spectrum' introduced below.

\subsection{The $q_{1}=q_{2}=\dots=q_{N}=q$ Case}


In the rainbow state model, the ground state is a tensor product of concentric maximally entangled singlets, or in the language of our model, $q_{i}=1$, for all $i$. Here, we show how the parameters of our model can be chosen such that all deformation parameters take the same value, $q_{i}=q$, for some chosen $q$ in the allowed range $0<q<\infty$. In this way, each concentric pair on our chain shares the same degree of pairwise entanglement, and all single-particle entanglement energies are equal.

We have defined $q_{1}=e^{\sinh^{-1}\left(\frac{h_{1}}{J_{1}}\right)}$ and $q_{i>1}=e^{\sinh^{-1}\left(\frac{\Tilde{h}_{i}}{\Tilde{J_{i}}}\right)}$, such that the condition $q_{1}=q_{i} \implies \frac{h_{1}}{J_{1}} = \frac{\Tilde{h}_{i}}{\Tilde{J_{i}}}$. Re-arranging equation \eqref{eq:q1} and setting $q_{1}=q$ yields
\begin{equation}
    \frac{h_{1}}{J_{1}} = \frac{1}{2}\left(q-\frac{1}{q} \right),
    \label{eq:h1J1_q}
\end{equation}
for some desired $q>0$. 

For all other pairs of sites the relations for the renormalised couplings must be used. For example. by dividing equation \eqref{eq:h2} by equation \eqref{eq:J2} and equating with \eqref{eq:h1J1_q} we obtain
\begin{equation}
	h_{2} = \frac{2J_{2}^{2}}{h_{1}} \frac{\left(1-q^{2}\right)^{2}}{\left(1+q^{2}\right)^{2}}.
\end{equation}
For any fixed value of $J_{2}$ this relation can be easily implemented to find the required transverse field parameter to produce some desired $q>0$.

In the same way, the ratio of equations \eqref{eq:ji'} and \eqref{eq:hi'} can be equated with \eqref{eq:h1J1_q} in order to obtain the general formula
\begin{equation}
	h_{i} = \frac{4J_{i}^{2}}{h_{i-1}} \frac{\left(1-q^{2}\right)^{2}}{\left(1+q^{2}\right)^{2}}, \hspace{0.2cm} i>2
\end{equation}

\begin{figure}
  \includegraphics[width=\linewidth]{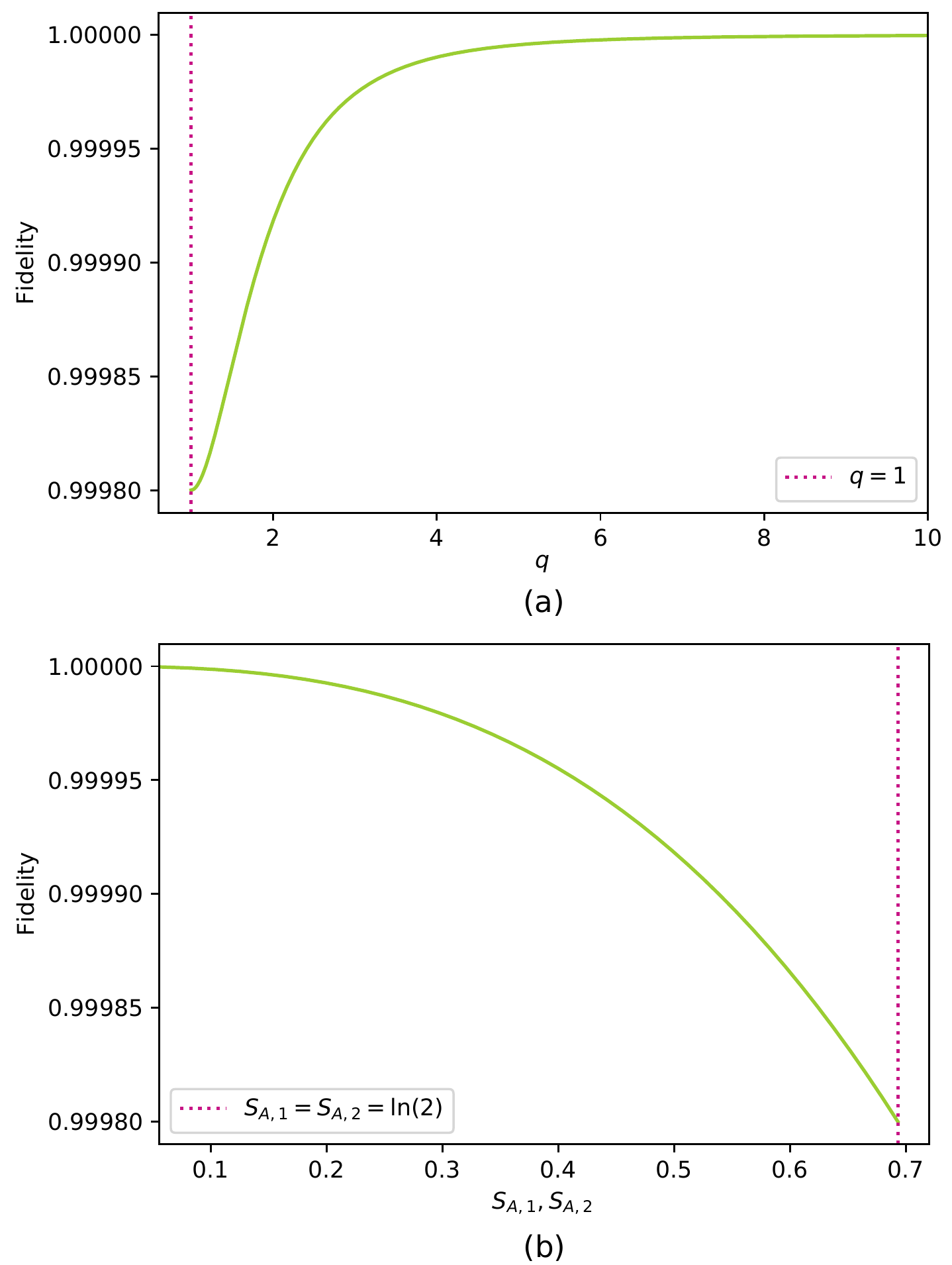}
\caption{(a) The variation of the ground state fidelity for $1\leq q\leq 10$ in the special case $q_{1}=q_{2}=q$ for $N=4$. Here $J_{1}=1, J_{2}=0.01$ (b) The corresponding variation of the fidelity with the associated range of equal von Neumann entanglement entropies across both pairs. All values in the range shown can be produced with a very high level of fidelity.
}
\label{fig:fid_q}
\end{figure}

By iterating through and systematically determining each successive value of the required transverse field for some fixed coupling profile, these relations allow us to produce a one-dimensional chain in which each concentric pair shares the same degree of pairwise entanglement. In Figures \ref{fig:fid_q}(a) and (b) it is shown how all values of $1\leq q\leq10$ and the corresponding entanglement entropies for the case $N=4$ can be produced with a very high level of fidelity. For just $J_{1}=1, J_{2}=0.01$ the error is of the order of $10^{-4}$.




\subsection{Prime Number Spectrum}

Prime numbers play an important role in number theory. The Fundamental Theorem of Arithmetics \cite{apostol1998introduction} states that every natural number greater than one can be uniquely factorised as a product of prime numbers
\begin{equation}
    N = 2^{n_{2}}3^{n_{3}}\dots p^{n_{p}}\dots,
\end{equation}
where $p$ is a prime and $n_{p}$ counts the number of times that $p$ appears in the factorisation of $N$. In this way, prime numbers can be thought of as the building blocks of all natural numbers.

Let us introduce the Moebius function, $\mu(n)$:
\begin{equation}
    \mu(n)=
    \begin{cases}
      1, &  n= 1, \\
      (-1)^{r}, & n=p_{1}\dots p_{r}, \\
      0, & \exists p, p^{2}\vert n,
    \end{cases}
\end{equation}
where $p$ are prime numbers. The symbol $p^{2}\vert n$ means that $p^{2}$ divides $n$. A square free integer is an integer whose factorization into products of primes does not contain any square of a prime numbers. $\mu(n)$ is therefore non-vanishing only on square free integers and its value is $+1$ if it contains an even number of primes and $-1$ if it contains an odd number of primes. In this way, the function $\mu$ is a sort of Fermi statistics if we think of the primes as being fermions.

Let us now consider an entanglement spectrum of the form
\begin{equation}
    \lambda_{k} = \frac{A_{F}\abs{\mu(k)}}{k^{s}}, \hspace{0.2cm} A_{F}, s>0, \hspace{0.1cm} k=1, \dots ,\infty.
\end{equation}
The normalization of the eigenvalues implies that
\begin{equation}
    1 = \sum_{k=1}^{\infty} \lambda_{k} = A_{F} \prod_{p} (1+p^{-s}) = A_{F} \frac{\zeta(s)}{\zeta(2s)},
\end{equation}
where we have used the Euler product formula
\begin{equation}
    \zeta(s) = \prod_{p} \frac{1}{1-p^{-s}}, \hspace{0.2cm} \text{Re}(s)>1.
\end{equation}
Using equation \eqref{eq:ES}, the entanglement energies for this spectrum are given by
\begin{equation}
    E_{k} = -\ln{\lambda_{k}} = -\ln{A_{F}} + s\ln{k}.
\end{equation}
Where $k$ is any square free integer. We equate this expression with that of the spectrum of a free fermionic system
\begin{equation}
    -\ln{A_{F}} + s\ln{k} = E_{0} + \sum_{i=1}^{N} n_{i}(k) \epsilon_{i}.
	\label{eq:primefreespectrum}
\end{equation}
If $k$ is a square free integer then from the Fundamental Theorem of Arithmetics one has
\begin{equation}
    k = 2^{n_{2}}3^{n_{3}}\dots p^{n_{p}}\dots, \hspace{0.3cm} n_{i}=0,1,
	\label{eq:k}
\end{equation}
such that taking the logarithm of \eqref{eq:k} yields
\begin{equation}
    \ln{k} = \sum_{p:prime} n_{p} \ln{p}, \hspace{0.3cm} n_{p}=0,1.
\end{equation}
Hence, equation \eqref{eq:primefreespectrum} is solved by
\begin{equation}
    E_{0} = -\ln{A_{F}} = \frac{\zeta(s)}{\zeta(2s)},
\end{equation}
\begin{equation}
    \epsilon_{p} = s\ln{p}.
	\label{eq:prime}
\end{equation}
The parameter $s$ can be thought of as an entanglement temperature since it is common to all eigenenergies. The relation \eqref{eq:prime} has also been considered in \cite{julia1990statistical, spector1990supersymmetry} with $\ln{p}$ being the single-particle energies of the primon gas. The partition function of this gas is related to the Riemann zeta function, $\zeta(s)$. In recent work, a prime number eigenvalue spectrum has also been experimentally realised by application of holographic optical traps \cite{cassettari2022holographic}, in agreement with previous theoretical results \cite{mussardo1997quantum}.

In Section \ref{sec:ordering}, we saw that the highest ground state fidelity is achieved by fixing our parameters $\{h_{i}\}$ and $\{J_{i}\}$ such that $\abs{\epsilon_{i}}\geq\abs{\epsilon_{i+1}}$. Thus, in order for our model to most accurately reproduce this prime number spectrum, we choose $\epsilon_{i}=s\ln{p_{i}}$ such that $p_{i}>p_{i+1}$. The required values of the set $\{h_{i}\}$, for some fixed coupling profile $\{J_{i}\}$, can then be simply read off from equations \eqref{eq:singpart1_}, \eqref{eq:i2} and \eqref{eq:ieven} in the Appendices.

\section{Conclusions}

In summary, we have introduced a spin chain that can produce arbitrary ground state free-particle correlations across a given bipartition. Our scheme is a generalisation of the rainbow states of concentric maximally entangled singlets to the case of concentric pairs, each one with arbitrary entanglement. The degree of entanglement is easily tuned by appropriately choosing the magnitude of local magnetic fields. The entanglement across the bipartition can be parametrised in terms of single particle energies of the entanglement spectra. We find that for a fixed coupling profile these energies are simple functions of the magnetic fields, thus providing direct accessibility and tunability.

To test the validity and applicability of our method we compare the fidelity of the predicted theoretical model with the exact diagonalisation of the spin system. The employed perturbation method has a symmetry in terms of the ordering of the concentric entangled states. By taking advantage of this symmetry we find the optimal order of magnetic fields that gives the best fidelities. Finally, we apply our method to two case scenarios. First, we consider the homogeneous case of concentric pairs with the same entanglement. Second, we consider the case of single particle energies of the entanglement spectra that are parametrised by prime numbers. This model is inspired by the similarity between the decomposition of free-system entanglement spectra in terms of single particle energies and the decomposition of integers in terms of prime numbers. In recent experimental work, holographic techniques have been developed allowing for the tuning of the energy spectrum of the single-particle Schr\"odinger equation \cite{cassettari2022holographic}. Notably a ‘prime number quantum potential’, $V_{N}(x)$, can be applied such that the single-particle Schr\"odinger equation has the lowest $N$ prime numbers as eigenvalues. These techniques demonstrate an interesting procedure for the reproduction of arbitrary sequences of integers as energy spectra with possible applicability to this work.

Our methodology can have a direct application in quantum technologies, whenever a very specific pattern of quantum correlations is required \cite{di2008nested, briegel1998quantum, franz2013quantum}. It can also simulate quantum phases of matter that require specific ground state correlations across a bipartition. Finally, our approach opens the way to investigate inhomogeneous spin chains in the presence of disordered magnetic fields, which is physically a common scenario, thus generalising previous approaches \cite{alkurtass2014optimal, langlett2022rainbow}.

\acknowledgements 
We would like to thank Gabriel Matos and Andrew Hallam for inspiring conversations. L.B. acknowledges support from EPSRC Grant No. EP/T517860/1. This work was also in part supported by EPSRC Grant No. EP/R020612/1. G.S. acknowledges financial support through the Spanish MINECO grant PID2021-127726NB-I00, the Comunidad de Madrid Grant No. S2018/TCS-4342, the Centro de Excelencia Severo Ochoa Program SEV-2016-0597 and the CSIC Research Platform on Quantum Technologies PTI-001.

\appendix

\section{Four Site Perturbation Theory}
\label{sec:4sitePT}
In order to illustrate the Real-Space RG approach, we apply perturbation theory to our Hamiltonian \eqref{eq:model} restricted to a chain of four sites
\begin{equation}
    H = H_{1} + \lambda V,
\end{equation}
where
\begin{equation}
H_{1} = J_{1}(\sigma_{-1}^{x}\sigma_{1}^{x}+\sigma_{-1}^{y}\sigma_{1}^{y})+h_{1}(\sigma_{-1}^{z}-\sigma_{1}^{z}),
\end{equation}
and
\begin{equation}
V = J_{2}^{\prime}(\sigma_{-2}^{x}\sigma_{-1}^{x}+\sigma_{-2}^{y}\sigma_{-1}^{y}+\sigma_{1}^{x}\sigma_{2}^{x}+\sigma_{1}^{y}\sigma_{2}^{y})+h_{2}^{\prime}(\sigma_{-2}^{z}-\sigma_{2}^{z}).
\end{equation}
Here the couplings $J_{2}=\lambda J_{2}^{\prime}$ and $h_{2}=\lambda h_{2}^{\prime}$, such that for any $\lambda\ll1$, the perturbative condition $J_{2},h_{2}\ll J_{1},h_{1}$ is ensured.

On two sites, $H_{1}$ has the following eigenstates
\begin{equation}
    \ket{\psi^{-}} = \frac{1}{\sqrt{[2]_{q_{1}}}} \left(q_{1}^{-\frac{1}{2}} \ket{\uparrow\downarrow} - q_{1}^{\frac{1}{2}}\ket{\downarrow\uparrow} \right),
\end{equation}
\begin{equation}
    \ket{\psi^{0}} = \ket{\uparrow\uparrow},
\end{equation}
\begin{equation}
    \ket{\psi^{1}} = \ket{\downarrow\downarrow},
\end{equation}
\begin{equation}
    \ket{\psi^{+}} = \frac{1}{\sqrt{[2]_{q_{1}}}} \left(q_{1}^{\frac{1}{2}} \ket{\uparrow\downarrow} + q_{1}^{-\frac{1}{2}}\ket{\downarrow\uparrow} \right),
\end{equation}
with eigenenergies $E_{1}=-[2]_{q_{1}}J_{1}, E_{s}=0, E_{t}=0$ and $E_{k}=+[2]_{q_{1}}J_{1}$ respectively, and $q_{1}$ as previously defined in equation \eqref{eq:q1}. 

When extended to a chain of four spins, the ground state subspace of $H_{1}$ becomes four-fold degenerate. We represent this subspace with the basis vectors:
$\{\ket{m}\} = \{\ket{m_{1}},\ket{m_{2}},\ket{m_{3}},\ket{m_{4}}\}$ $=\{ \ket{\uparrow}_{-2}\ket{\psi^{-}}_{-1,1}\ket{\uparrow}_{2}$, $\ket{\uparrow}_{-2}\ket{\psi^{-}}_{-1,1}\ket{\downarrow}_{2}$, $\ket{\downarrow}_{-2}\ket{\psi^{-}}_{-1,1}\ket{\uparrow}_{2}, $ $\ket{\downarrow}_{-2}\ket{\psi^{-}}_{-1,1}\ket{\downarrow}_{2} \}$.
The first-order corrections arise due to the action of the perturbative term on the ground state subspace. This is quantified via the computation of the matrix elements of the effective Hamiltonian to first order
\begin{equation}
    H_{\alpha,\beta}^{eff \left(1\right)} = \bra{m_{\alpha}}V\ket{m_{\beta}},
\end{equation}
yielding
\begin{equation}
    H^{eff (1)} = 2h_{2}^{\prime}\begin{pmatrix}
    0 & 0 & 0 & 0\\
    0 & 1 & 0 & 0 \\
    0 & 0 & -1 & 0 \\
    0 & 0 & 0 & 0
    \end{pmatrix} 
\end{equation}
in the basis $\{\ket{m}\}$. By inspection, it can be seen that the first-order effective Hamiltonian term is therefore $H^{eff (1)}= h_{2}^{\prime}(\sigma_{-2}^{z}-\sigma_{2}^{z})$. The first-order ground state energy correction is found by diagonalizing the above matrix. It is clear that the degeneracy is only partially lifted to first order. It is therefore necessary to consider the second order corrections that arise due to the overlap with states from each of the excited state subspaces. These excited state subspaces are found in the same way as the set $\{\ket{m}\}$ by taking the tensor product of the two-qubit computational basis with the excited eigenstates of $H_{0}$: $\{\ket{s}\}=\{\ket{s_{1}},\ket{s_{2}},\ket{s_{3}},\ket{s_{4}}\}$ $=\{\ket{\uparrow\uparrow\uparrow\uparrow}$, $\ket{\uparrow\uparrow\uparrow\downarrow}$, $\ket{\downarrow\uparrow\uparrow\uparrow}$, $\ket{\downarrow\uparrow\uparrow\downarrow}\}$, $\{\ket{t}\}=\{\ket{t_{1}},\ket{t_{2}},\ket{t_{3}},\ket{t_{4}}\}$ $=\{\ket{\uparrow\downarrow\downarrow\uparrow}$, $\ket{\uparrow\downarrow\downarrow\downarrow}$, $\ket{\downarrow\downarrow\downarrow\uparrow}$, $\ket{\downarrow\downarrow\downarrow\downarrow}\}$, $\{\ket{k}\} = \{\ket{k_{1}},\ket{k_{2}},\ket{k_{3}},\ket{k_{4}}\}$ $= \{ \ket{\uparrow}_{-2}\ket{\psi^{+}}_{-1,1}\ket{\uparrow}_{2}$, $\ket{\uparrow}_{-2}\ket{\psi^{+}}_{-1,1}\ket{\downarrow}_{2}$, $\ket{\downarrow}_{-2}\ket{\psi^{+}}_{-1,1}\ket{\uparrow}_{2}$, $\ket{\downarrow}_{-2}\ket{\psi^{+}}_{-1,1}\ket{\downarrow}_{2} \} \}$. Such that the full set of excited states, $\{\ket{n}\} = \{\{\ket{s}\}, \{\ket{t}\}, \{\ket{k}\} \}$.

The matrix elements of the effective Hamiltonian to second-order are found from
\begin{equation}
    H_{\alpha,\beta}^{eff (2)} = \sum_{i=1} \frac{\bra{m_{\alpha}}V\ket{n_{i}}\bra{n_{i}}V\ket{m_{\beta}}}{E_{-}-E_{i}}.
\end{equation}
The computation of which yields
\begin{equation}
    H^{eff (2)} = \frac{(2J_{2}^{\prime})^{2}}{(1+q_{1}^{2})E_{1}}\begin{pmatrix}
    1+q_{1}^{2} & 0 & 0 & 0\\
    0 & 2q_{1}^{2} & -2q_{1} & 0 \\
    0 & -2q_{1} & 2 & 0 \\
    0 & 0 & 0 & 1+q_{1}^{2}
    \end{pmatrix}.
\end{equation}
Combining our first and second-order perturbative terms we derive an expression for the effective Hamiltonian correct to $\mathcal{O}(\lambda^{2})$
\begin{align}
    H^{eff} &\approx E_{1}\mathbf{1}_{4} + \lambda H^{eff (1)} + \lambda^{2}H^{eff (2)} \\
    &= \left(E_{1}+\frac{(2\lambda J_{2}^{\prime})^{2}}{E_{1}}\right)\mathbf{1}_{4} \\
& \qquad - \frac{q_{1}(2\lambda J_{2}^{\prime})^{2}}{(1+q_{1}^{2})E_{1}}(\sigma_{-2}^{x}\sigma_{2}^{x}+\sigma_{-2}^{y}\sigma_{2}^{y}) \nonumber \\
& \qquad +\left(\lambda h_{2}^{\prime}-\frac{2(1-q_{1}^{2})(\lambda J_{2}^{\prime})^{2}}{(1+q_{1}^{2})E_{1}}\right)(\sigma_{-2}^{z}-\sigma_{2}^{z}) \nonumber\\
    &= \mathcal{C}\mathbf{1}_{4} +\Tilde{J_{2}}(\sigma_{-2}^{x}\sigma_{2}^{x}+\sigma_{-2}^{y}\sigma_{2}^{y})+\Tilde{h}_{2}(\sigma_{-2}^{z}-\sigma_{2}^{z})
\end{align} 
In this way we obtain the following expressions for our renormalized parameters
\begin{equation}
    \Tilde{J_{2}} = \frac{4J_{2}^{2}}{[2]_{q_{1}}^{2}J_{1}},
\end{equation}
\begin{equation}
	\Tilde{h}_{2} = h_{2} - \frac{2\left(q_{1}-\frac{1}{q_{1}}\right)J_{2}^{2}}{[2]_{q_{1}}^{2}J_{1}}.
\end{equation}
Note that, in the case $h_{1}=h_{2}=0$ such that $q_{1}=q_{2}=1$, $\Tilde{h}_{2}$ vanishes and $\Tilde{J_{2}}$ returns to that of the inhomogeneous XX model re-scaling as seen in equation \eqref{eq:XXrescaling} as expected.

Diagonalization of the effective Hamiltonian yields the ground state
\begin{align}
    \ket{\psi} &= \frac{1}{\sqrt{[2]_{q_{2}}}} \left(q_{2}^{-\frac{1}{2}} \ket{m_{2}} - q_{2}^{\frac{1}{2}}\ket{m_{3}} \right) \\
    &= \left(\frac{1}{\sqrt{[2]_{q_{1}}}}(q_{1}^{-\frac{1}{2}}\ket{\uparrow\downarrow}_{-1,1}-q_{1}^{\frac{1}{2}}\ket{\downarrow\uparrow}_{-1,1}) \right) \\
& \qquad \otimes\left(\frac{1}{\sqrt{[2]_{q_{2}}}}(q_{2}^{-\frac{1}{2}}\ket{\uparrow\downarrow}_{-2,2}-q_{2}^{\frac{1}{2}}\ket{\downarrow\uparrow}_{-2,2}) \right) \nonumber
\end{align}
with corresponding ground state energy
\begin{equation}
    E_{2} = E_{1}-\frac{4J_{2}^{2}}{[2]_{q_{1}}J_{1}} +\Tilde{E_{2}},
\end{equation}
where
\begin{equation}
	\Tilde{E_{2}} = -[2]_{q_{2}}\Tilde{J_{2}}.
\end{equation}

\section{Relationship Between Real Model Parameters and Single-Particle Entanglement Energies}
\label{sec:entenergies_app}

In Section \ref{sec:entspec} the following relationship was found between the renormalised coupling parameters and single-particle entanglement energies
\begin{equation}
	\frac{\Tilde{h}_{i}}{\Tilde{J_{i}}} = -\sinh{\left(\frac{\epsilon_{i}}{2}\right)}. \nonumber
\end{equation}

By combining this relation with equations \eqref{eq:ji'} and \eqref{eq:hi'}, we can obtain expressions that directly relate the desired single-particle entanglement energies to the required ratio of real parameters in our model. The recursive nature of the formulae for the re-scaled parameters means that in order to engineer some single-particle energy $\epsilon_{i}$ it is necessary to have previously established some value for all previous $i-1$ energies.

In the case $J_{1}, h_{1} \gg J_{i\neq1},h_{i\neq1}$ that we consider, the central terms $J_{1}$ and $h_{1}$ do not get re-scaled, therefore simply
\begin{equation}
    h_{1} = -J_{1}\sinh{\left(\frac{\epsilon_{1}}{2}\right)},
    \label{eq:singpart1_}
\end{equation}
for some desired single-particle energy, $\epsilon_{1}$, and fixed value of the central coupling term.
In Appendix \ref{sec:4sitePT} we have found exact expressions for the renormalized parameters $\Tilde{J_{2}}$ and $\Tilde{h}_{2}$. By substituting these into \eqref{eq:singpart} and fixing all other parameters, we obtain an expression for the required transverse field
\begin{equation}
    h_{2} = -\frac{J_{2}^{2}}{\cosh^{2}{\left(\frac{\epsilon_{1}}{2}\right)}J_{1}}\left[\sinh{\left(\frac{\epsilon_{2}}{2}\right)}+\sinh{\left(\frac{\epsilon_{1}}{2}\right)} \right].
	\label{eq:i2}
\end{equation}
 Repated iterations of this process yield a general form for the required transverse field parameter for any pair of sites $-i$ and $i$, when $i>2$
\begin{multline}
    h_{i>2} = \frac{J_{i}^{2}}{\cosh^{2}\left(\frac{\epsilon_{i-1}}{2}\right)h_{i-1}}\left[\sinh{\left(\frac{\epsilon_{i}}{2}\right)}+\sinh{\left(\frac{\epsilon_{i-1}}{2}\right)} \right] \\ \times
\left[\sinh{\left(\frac{\epsilon_{i-1}}{2}\right)}+\sinh{\left(\frac{\epsilon_{i-2}}{2}\right)} \right].
\label{eq:ieven}
\end{multline}
Thus, it is made explicit how the real parameters of our model can be selected such as to produce any desired set of single-particle entanglement energies.

\bibliographystyle{apsrev4-2}
\bibliography{IntDistPri3101}

\end{document}